\documentclass[pre,aps,nofootinbib,twocolumn,showpacs]{revtex4}
\usepackage{graphics}
\usepackage{epsfig}

\begin{document}
\newcommand{\be}{\begin{equation}}
\newcommand{\ee}{\end{equation}}

\title{Study of the multi-species annihilating random walk transition at zero
branching rate - cluster scaling behavior in a spin model
\footnote{To appear in Phys. Rev. E in Nov. 2003}}

\author{N\'ora Menyh\'{a}rd$^1$ and G\'eza \'Odor$^2$}
\affiliation{$^1$ Research Institute for Solid State
Physics and Optics, H-1525 Budapest, P.O.Box 49, Hungary \\
$^2$ Research Institute for Technical Physics
and Materials Science, H-1525 Budapest, P.O.Box 49, Hungary}

\begin{abstract}

Numerical and theoretical studies of a one-dimensional spin model with 
locally broken spin symmetry are presented.
The  multi-species annihilating random walk transition found at zero 
branching rate previously is investigated
now concerning the cluster behaviour of the underlying
spins. Generic power law behaviors are found,
besides the phase transition point, also in the
active phase with fulfillment of the hyperscaling law.
On the other hand  scaling laws connecting bulk- and cluster
exponents are broken  - a possibility in no contradiction with basic
scaling assumptions because of the missing absorbing phase.
\end{abstract}
\maketitle

\section{ Introduction }
The study of non-equilibrium model systems has attracted great attention
in recent years. A variety of phase transitions have been found
characterized by critical exponents, both static and dynamic. 
Of special interest are transitions from a fluctuating active state into 
an absorbing one.
A wide range of models with transitions into absorbing states was
found  to belong to the directed percolation (DP) universality
class \cite{DP}. Another universality class of interest is the so-called
parity conserving (PC) class \cite{jen94,dani,kim94,park}.
The mostly  studied particle model in this class is branching
annihilating random walk (BARW) in one dimension (1d) with an even 
number of offsprings ($2A\rightarrow 0$, $A\rightarrow 3A$, in the simplest 
case). The first example of model systems exhibiting PC-type transition was 
given, however, in two 1d cellular automata  by Grassberger \cite{gra8489}.
The prototype {\it spin}-model for PC-type phase transitions was
proposed by one of the authors \cite{men94} by introducing a class
of non-equilibrium kinetic Ising models (NEKIM) with combined
spin-flip dynamics \cite{gla63} at zero temperature, $T=0$ and
Kawasaki spin-exchange kinetics \cite{kaw72} at $T=\infty$.

Transitions between active and absorbing phases have been, however, 
mostly studied in particle-type
models. The N-BARW2 model is a classical stochastic system of N types
of particles with branching annihilating random walk. 
For $N>1$ $N$ types of particles $A_i$ perform diffusion, pairwise
annihilation of the same species and branching $A_i \rightarrow A_i+2A_j$
with rate $\sigma$ for $i=j$ and with rate $\sigma'/{(N-1)}$ for $i\neq j$.
 According to field 
theory \cite{Card97} in this model the rate $\sigma$ flows to zero
under coarse-graining renormalization which implies that 
the model is always active
except for the annihilation fixed point at
$\sigma^{'}=0$. It forms a  universality class, the so called N-BARW2,
different from DP and PC, with  well
known bulk critical exponents in 1d.

In the NEKIM model a global asymmetry of the spins (magnetic field)
is known to change the  PC transition into the DP type.
 \cite{parkh,meod96}. The question arises what 
is the effect of a {\it local}  breaking of the spin symmetry in such a spin system.
The first indication in this direction has come from a work of Majumdar et al.
\cite{mdg} who studied the coarsening dynamics of of a Glauber-Ising chain
with strong asymmetry
in the annihilation rate maximally favouring '-' spins (MDG-model). 
These authors found the result that while the '+' domains still coarsen
as $t^{1/2}$, the '-' domains coarsen slightly faster as $t^{1/2}\log (t)$.
As a result at  late times,  the system started from a random initial state
decays into a fully compact state where all spins become  '-' in a slow
logarithmic way $1/{\log(t)}$.

In 
a previous paper \cite{mo02} the  authors  presented an 
asymmetric spin-model (NEKIMA)
by generalizing the NEKIM model which includes as a special case the
MDG model. In NEKIMA  there is local spin-asymmetry both in the annihilation
rate (favouring '-' spins) and the diffusion-like  spin-flip rate 
(favoring '+' spins) and thus acting oppositely.
Global scaling properties of the model have been 
investigated numerically as
well as using cluster MF approximation. 
The N-BARW2
transition, for which no spin model had been known previously,
 was found at zero spin-exchange rate.
 In the present paper we further investigate this model
 at such parameter values for which in the original NEKIM
 model PC-type transition takes place. In the plane of the
 spin-asymmetry parameter and kink-branching probability
 we have found, by computer simulations as well as by cluster mean-field
calculations, a phase diagram showing a reentrant directed-percolation  
line. 
 Our main purpose, however, has been to investigate the cluster
 development properties a) at and in the vicinity of the N-BARW2
 line and b). in the rest of the parameter space considered.
 For the mean population size $n(t)\sim t^\eta$, for the mean square
 spreading of spins $ R^{2}(t) \sim t^z$ and for the survival probability
 $P(t)\sim t^{-\delta}$
 generic scaling behaviour has been found via computer simulations 
in (almost) the whole plane of 
the phase diagram with fulfillment of the hyperscaling law.
 Upon crossing the line of  zero branching rate  (where the phase transition
 takes place), however, dynamic scaling is
 found to be violated concerning laws connecting bulk exponents and 
 cluster ones. We trace back such a possibility
to the  circumstance that  the  absorbing
phase is missing by the N-BARW2 transition.

\section{The model and previous results}

The general form of the Glauber spin-flip transition rate in
one-dimension for spin $s_i$ sitting at site $i$ is  \cite{gla63} 
($s_i=\pm1$):
\begin{equation}
w(s_{i},s_{i-1},s_{i+1}) ={\Gamma\over{2}}(1+\tilde\delta s_{i-1}s_{i+1})
[1 - {1\over2}s_i(s_{i-1} + s_{i+1})]
\label{Gla}
\end{equation}
at zero temperature. (Usually the Glauber model is understood as
the special case  $\tilde\delta=0$, $\Gamma=1$.)

The kink $\rightarrow 3$ kink processes are introduced  via
the exchange rate 
\begin{equation}
w_{ex}(s_i,s_{i+1})={ p_{ex}\over{2}}(1-s_{i}s_{i+1})
\label{Kaw}
\end{equation}

This model (called NEKIM), for 
 negative values of $\tilde\delta$ in eq.(1) shows a line of PC-transitions
in the plane of the parameters ($p_{ex}$,$\tilde\delta$) 
 \cite{men94}.
In NEKIMA \cite{mo02} the authors have extended
the  above model by introducing local symmetry breaking in
the  spin-flip rates 
of the  $+$ and $-$ spins as follows. Concerning the annihilation rates
the prescription in \cite{mdg} is followed:
\begin{equation}
w(+;--)=1, \,\,  
w(-;++)=0,
\label{mdg}
\end{equation}
while further   spin symmetry breaking is introduced
in the diffusion part of the Glauber transition rate as follows.
In calculating the transition rates
\begin{equation}
p\equiv w(-;+-)=w(-;-+)=\Gamma /2 (1-\tilde\delta)
\end{equation}
the Glauber form,  eq.(\ref{Gla}), is used unchanged, while
 $w(+;+-)$  and  $w(+;-+)$ are allowed,  to take smaller values :
\begin{equation}
p_+\equiv w(+;+-)=w(+;-+)\leq p.
\label{p+}
\end{equation}
In this way , by locally favoring the $+$ spins, the effect of the 
other dynamically induced field  arising from the prescription
(eq.\ref{mdg}) is counterbalanced.
The spin-exchange part of the NEKIM model remains unchanged, eq.(\ref{Kaw}). 
  It is worth mentioning that the lifting of the strong
restriction in eq.(\ref{mdg}) together with applying  spin anisotropy ${p_+}<
\Gamma/2 (1-\delta) $ has the same effect as a global magnetic field favoring
$+$ spins.

\begin{center}
\begin{figure}
\vspace{4mm}
\epsfxsize=80mm
\epsffile{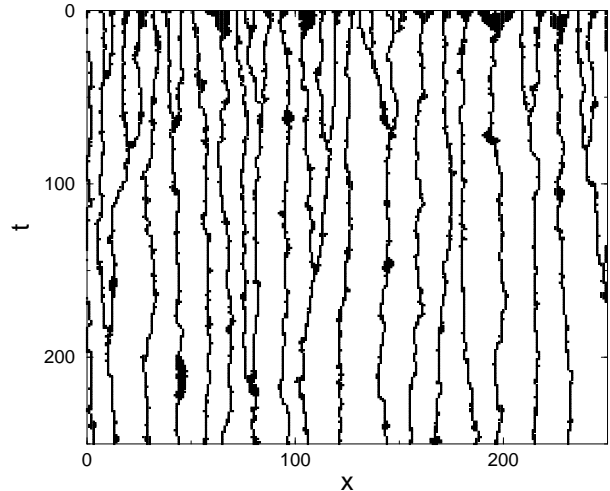}
\caption{Space-time development of '+' (white) and '-' (black) spins 
evolving from a random initial state for $p_+=.1$, $p_{ex}=0$.
Throughout the whole paper $t$ is measured in units of Monte-Carlo sweeps.}
\label{fig1}
\end{figure}
\end{center}

For $p_{ex}=0$, the absorbing states in the extreme situation 
$p_{+}=0$ when diffusion-like spin- flipping
maximally favors $+$ spins, are states with single frozen $-$ spins like 
$+-+++-++-+++$. By increasing $p_{+}$ from zero, a slow random walk of these 
lonely $-$- spins starts and by annihilating random walk only one of them 
survives and performs RW (see Fig.\ref{fig1}) . 
The all $+$ and all $-$ states are, of course, also absorbing.

In \cite{mo02} the authors have studied the following  global quantities 
for different values of $p_{+}<p$ :
the density of kinks  
 as a function of time, 
starting from a 
random initial distribution of spins for $p_{ex}=0$
\begin{equation}
\rho(t) \sim t^{-\alpha}
\label{alpha}
\end{equation}
and its asymptotic values for finite but small values of $p_{ex}$
\begin{equation}
\rho_\infty (p_{ex})\sim {p_{ex}}^\beta .
\label{beta}
\end{equation}
The  results obtained within error of simulations, 
$\alpha= .5$  and $\beta = 1.0$,
pointed to the presence of a N-BARW2 transition.
 Finite size scaling 
behaviour was also examined to find the other two bulk exponents,
those of
$\xi$ , the correlation length and
$\tau$,  the characteristic time:

\begin{equation}
\xi \sim p_{ex}^{-\nu_{\bot}} , \tau\sim \xi^Z
\label{ksi}
\end{equation}
where $Z$ is the dynamical critical exponent.
The expectation for a N-BARW2 transition at zero
branching rate was justified by the values :
$\nu_{\bot}=1.0$ and $Z=2.0$, which were  found within error of simulations. 
We also found the expected phase diagram of a line
of DP transitions in the $(p_{+}, p_{ex})$ plane (instead of the PC-line
of NEKIM).
\vglue .5cm

\section{Cluster behavior at and below the N-BARW2 transition}

Spreading from a localized source at criticality 
 is usually described by the following three quantities
\begin{equation}
P(t)         \sim t^{-\delta},\\ 
{n(t)}    \sim t^{\eta}, \\
{R^2(t)}  \sim t^z.        
\label{clex}
\end{equation}
where $n(t)$ denotes the mean population size, $R^2(t)$ is the
mean square spreading of particles (here spins) about the origin and
$P(t)$ is the survival probability.
In most cases these quantities are  defined for particles, in the present case,
however,
like for studying compact directed percolation of an Ising chain
\cite{DiTre}
they will be used for spins.

In the active phase the survival probability defines a further
useful critical exponent $\beta^{'}$, ($ \nu_{\parallel}=Z\nu_\bot$ ) 
\begin{equation}
P \sim t^{-\delta} g(p_{ex}t^{1/{\nu_{\|}}})
\end{equation} 
as
\begin{equation}
P_\infty \sim {p_{ex}}^{\beta^{'}}, \ \
\beta^{'} =\nu_{\|}\delta
\end{equation}
In the present case the following parameter values of NEKIMA 
were used in the simulations: $\Gamma =.5$, $\tilde\delta=-.565$ 
($p=.39125$).
 The phase diagram in the ($\Delta\equiv\frac{p-p_+}{p}$, $p_{ex}$)
 plane is shown on Fig.2.
The origin $(0,0)$ will be called "MDG-point" as at this point
$p_+ = p$ and the model is the same as treated in \cite {mdg}(though
the values of $\tilde\delta$, and $\Gamma$ 
are different). The line $p_{ex}=0$ is a line of compactness,
 as will be discussed in the
next  section.
Also other details of the phase diagram will be explained
later.

\begin{center}

\begin{figure}
\epsfxsize=80mm
\epsffile{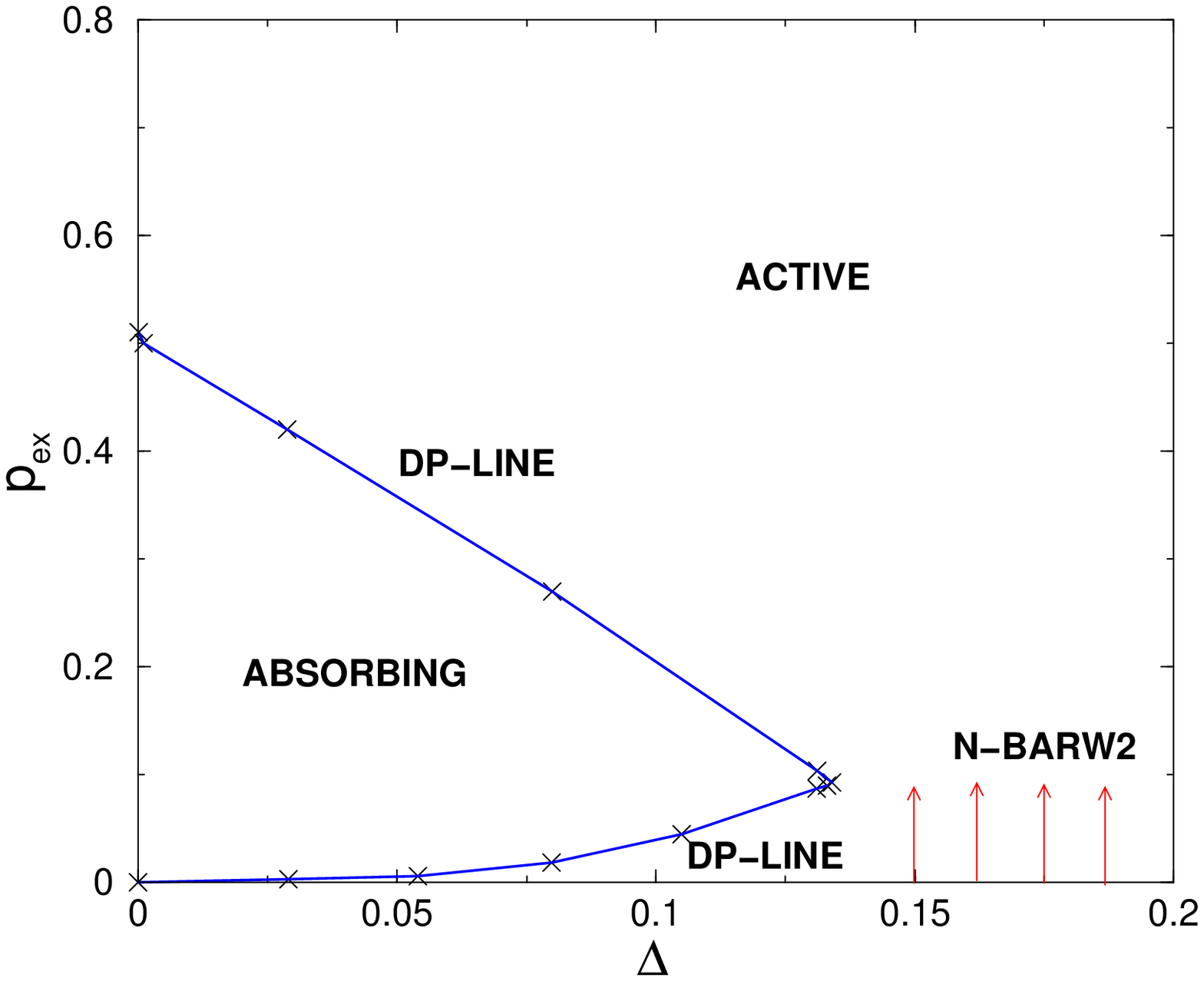}
\vspace{4mm}
\caption{Phase diagram of the NEKIMA model for $\delta=-.565$,
 $\Gamma=.5$. The absorbing phase is fully '$-$'.}
\label{phasedia}
\end{figure}
\end{center}
In NEKIMA '+' and '-' spins are not symmetric, therefore we
have investigated two kinds of clusters. 
Namely, the development
of the '-' cluster-seed 
 was started from a wholly '+' environment
at t=0: ++++++++ - ++++++++, while the '+' cluster
from a sea of '-' spins: - - - - - - + - - - - -  .
We will call them '-' cluster and '+' cluster, respectively.
The simulations have been performed with several values of
$p_+$ and $p_{ex}$; for $t_{max}=5.10^3$ MC steps and for averages
over $10^4$ samples. The local slopes
\begin{equation}
-\delta(t)={\ln[P(t)/P(t/m)]\over\ln m}
\label{locexp} 
\end{equation}
(and similarly for $\eta(t)$ and $z(t)$) as a function of $1/t$ are
plotted, as usual in case of simulations for critically behaving
quantities. In (\ref{locexp}) $m>1$ is an arbitrary factor
which we took to be equal to 5. The results obtained in different regions
of the phase diagram, Fig.\ref{phasedia}, are summarised on  
Table I and Table II. As it is apparent from Table I, the '+' cluster
does not change its exponents by crossing the $p_{ex}=0$ line. The 
'-' cluster's exponents, however, change abruptly.

\begin{center}
\begin{figure}
\epsfxsize=80mm
\epsffile{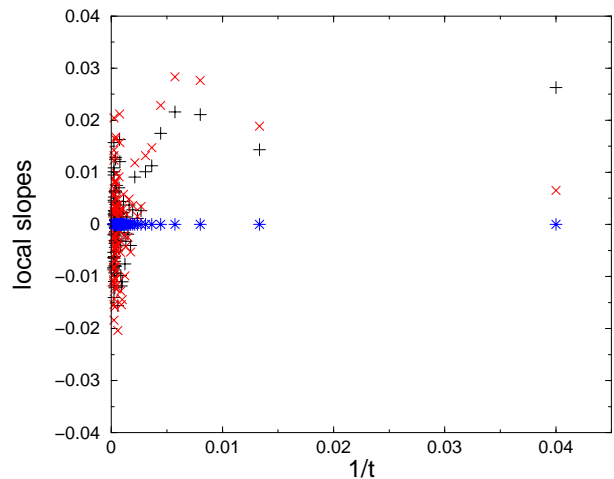}
\vspace{4mm}
\caption{Cluster exponents for '-' cluster at $p_{+}=.3$, $p_{ex}=0$
Number of MC steps: $10^5$, number of averages: $10^3$.
"+" correspond. to $\eta$, "x"-s to $z/2$, *-s to $\delta$}
\label{ncl}
\end{figure}
\end{center}

\begin{center}
\begin{figure}
\epsfxsize=80mm
\epsffile{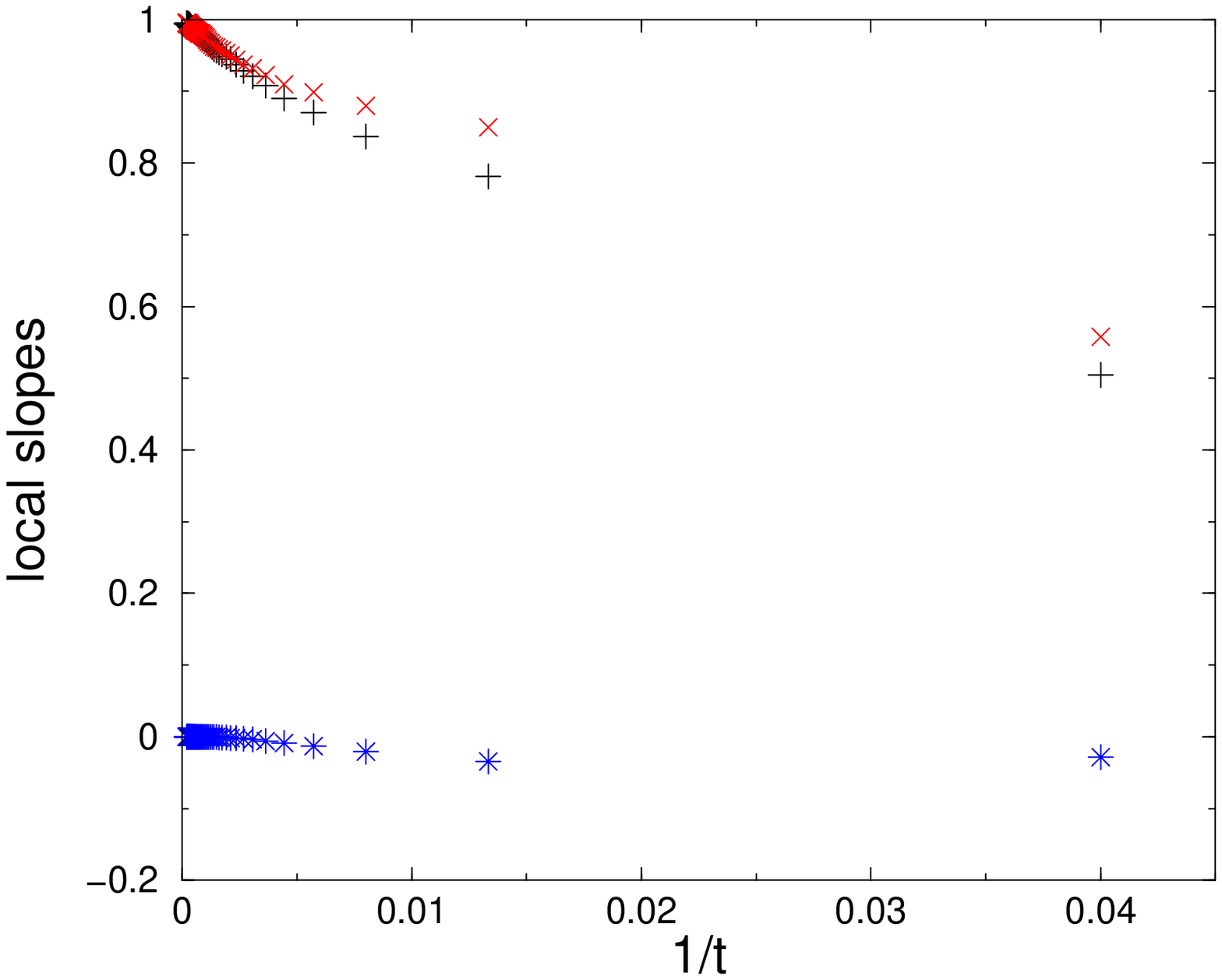}
\vspace{4mm}
\caption{Cluster exponents for "-" cluster at $p_{+} = .3$, $p_{ex}=0$
Number of MC steps: $10^5$, number of averages: $10^3$. 
"+"-es correspond to $\eta$,  "x"-s to $z/2$, *-s to $\delta$}
\label{ncl2}
\end{figure}
\end{center}

\begin{center}
\begin{figure}
\epsfxsize=80mm
\epsffile{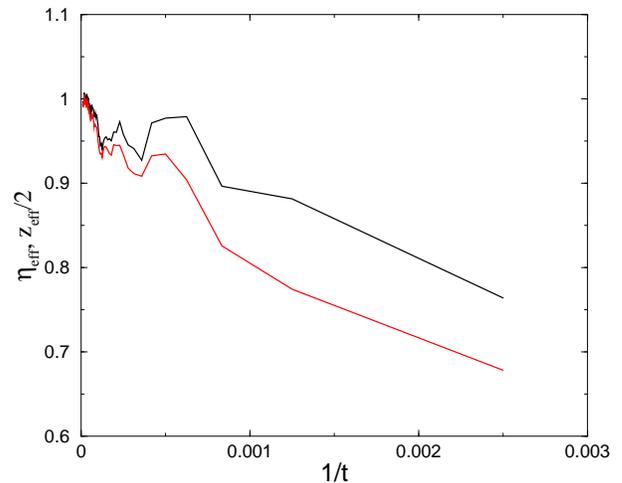}
\vspace{4mm}
\caption{Cluster spreading exponents for '+' cluster at parameter
values $p_+=.3$,$p_{ex}=.02$ ( $\delta=-.565$, $\Gamma=.5$.)
 Number of MC
steps: $10^5$, number of averages: $10^3$. Upper curve:$\eta_{eff}$,
lower curve $z_{eff}/2$.}
\label{plcluslong}
\end{figure}
\end{center}

For the case $p_{+}=.3$  Figs.\ref{ncl} and \ref{ncl2} shows the local 
exponent values, (eq.\ref{locexp}), for $p_{ex}=0$ i.e. at the N-BARW2 
transition point and for $p_{ex}=.02$, i.e. in the active phase.
Here and in most cases of our simulations the number
of MC steps has been $5\times 10^3$ with averaging over $2\times 10^4$
different runs. In some cases , however, much longer runs
have also been carried out up to $10^5$ MC steps
to corroborate these results, see Fig.\ref{plcluslong}. 

\begin{table*}
\begin{ruledtabular}
\caption{Cluster critical exponents at and near the N-BARW2 transition
point. The hyperscaling law $\eta+\delta=z/2$ (see Section V.)
is satisfied.}
\begin{tabular}{|c|c|c|c|c|} 
 exponents & $p_{ex} = 0$, '+' &$p_{ex} = 0$ '-'& $p_{ex}\neq 0$ '+'& 
$p_{ex}\neq 0$ '-'   \\ \hline
$\eta$ &1. 0 & 0.0 & 1.0 & 1.0 \\
$\delta $  & 0.0 & 0.0  & 0.0& 0.0  \\
$ z  $   & 2.0 & 0.0 & 2.0 & 2.0 \\ 
\end{tabular}
\end{ruledtabular}
\end{table*}

\begin{table*}
\begin{ruledtabular}
\caption {Cluster critical exponents at and near the DP-line. 
For abbreviations see Fig.2. The hyperscaling-law, valid for DP transitions, 
$\eta+2\delta = z/2$ (see Section V) is satisfied}
\begin{tabular}{|c|c|c|c|c|c|c|} 
 exponents & on DP-line '+'& on DP-line '-'&ABSO-phase '+'&ABSO-phase '-'& ACTIVE-phase '+'&ACTIVE-phase '-"  \\ \hline
$\eta$ &.31 & 1.0 & exponential& 1.0& 1.0& 1.0\\
$\delta$ & .16 & .0& exponential & 0.0  & 0.0  &  0.0  \\
$ z$ & 1.26 &  2.0& exponential & 2.0 &  2.0 &  2.0 \\ 
\end{tabular}
\end{ruledtabular}
\end{table*}

For comparison let us recall the well-known values for the above
exponents in case of the compact directed percolation point
of the Domany-Kinzel cellular automaton \cite{DoKi}.
Dickman and Tretyakov \cite{DiTre} have given the results in this
context as follows: $\eta=0$, $\delta=1/2$ and $z=1$.
( The same as for the Glauber-Ising model at $\tilde\delta=0$,
$\Gamma=1.0$). It is of some interest to present the measured
cluster exponents at the origin of the phase diagram, Fig.2,
which is the (equivalent of the)MDG-point. Here we found for the
'+' cluster: $\eta=0$, $\delta=1/2$, $z=1$ 
while for the '-' cluster: $\eta=1/2$, $\delta=0$, $z=1$
(with the same accuracy
as most of our results here ($ t_{max}=5.10^3$ MC steps)).
 These data are summarized on
TABLE III.
Because of the relatively low upper limit in time of
most of our simulations as given above,
 the possibility of the presence of a $\log(t)$
correction at the MDG point can not be excluded.

\begin{table*}
\begin{ruledtabular}
\caption{ Cluster critical exponents in case of the Glauber-Ising and
MDG parameter values. The hyperscaling law $\eta+\delta = z/2$
is satisfied.}
\begin{tabular}{|c|c|c|c|c|}
  exponents     & Glauber-Ising , '+' & Glauber-Ising,  '-'& 
MDG,  '+'&
 MDG  '-'   \\ \hline
$\eta$ & 0 & 0 & 0.0 & 0.5 \\
$\delta $  & 1/2 & 1/2  & 0.5& 0.0  \\
$ z  $   & 1 & 1 & 1.0 & 1.0 \\
\end{tabular}
\end{ruledtabular}
\end{table*}

\section{Breaking of a scaling law}

According to the previous section the result for the critical exponent
of the mean square distance of spreading from the origin, $z$,
is equal to $2.0$ within error of numerical simulations.
For the dynamical critical exponent the value $Z=2.0$ was obtained,
in the whole regime ($p_+$ values ) of the N-BARW2 transition. 

On the other hand, at BARW-type transitions, like
DP and PC transitions, 
the following scaling law
connects the above two critical exponents:
\begin{equation}
z=2/Z
\label{zZ}
\end{equation}
This relation is usually  quoted as a consequence of dynamical
scaling.
Using the above cited results, however, eq.(\ref{zZ}) is broken.
The  possibility of 
breaking this scaling law is actually
 due to the circumstance, that the N-BARW2 transition
point lies at the zero value of the branching probability, $p_{ex}=0$ 
and there is no absorbing phase,
with exponentially decreasing space- and time dependences.
To support this point let us recall the way Mendes et al. 
\cite {MDHM} derived the relation (\ref{zZ}).

They started from the general expression for the density of
particles(kinks) at space-point $r$ in the  absorbing
phase $\Delta < 0$ ( here $\Delta$ denotes the deviation
 from the critical point)and at large fixed value of 
$t$ (for $d=1$):
\begin{equation}
\rho(r,t) = t^{\eta-z/2}F({r^2}/{t^{z}}, \Delta {t^{1/{\nu_{\|}}}})
\label{rho}
\end{equation}
In the absorbing phase the function $\rho(r,t)$ is expected to decrease
exponentially as $\rho(r,t) \sim \exp{-r/{\xi}}$, where
$\xi \sim \Delta^{-\nu_{\bot}}$. This form implies for $F(u,v)$
 (with $v<0 $) the form
\begin{equation}
F(u,v) \rightarrow \exp{(-C\sqrt{u}|v|^{\nu_{\bot}})} 
\end{equation}

where $C>0$ is  constant. 
For $\xi$ to be time-independent the scaling law is required:

\begin{equation}
z= \frac{2\nu_{\bot}}{\nu_{\parallel}} = \frac{2}{Z}
\label{znu}
\end{equation}
This  scaling law which is not  fulfilled
in the presently discussed model.
Moreover, the bulk quantity, the time dependent kink density
\begin{equation}
\rho(t) \sim t^{-\alpha}
\end{equation}
and the expression obtainable from eq.(\ref{rho})
\begin{equation}
\rho (t)\sim t^{\eta- z/2}
\label{etaz}
\end{equation}
are also in conflict. Namely, while all the simulations have resulted
in $ \alpha =.5$ within error and this is in agreement with
the scaling law $ \alpha = \frac{\beta}{{\nu_{\|}}}$, 
according to the values given in Table I the exponent in eq.(\ref{etaz})
is zero, again within the error of simulations.
(It is to be noted, that this conflict is no more present concerning
the exponent values at $p_{ex}\neq 0$, where $\alpha=0$ and
$\eta -z/2 =0$,  as well.) The apparent contradiction, however, is
resolved again by recalling that cluster exponents and bulk-exponents
are allowed to be not  connected by a scaling law. 


\section {Hyperscaling}
The generalized hyperscaling law \cite{MDHM}
 was developed for systems with
multiple absorbing configurations  and reads:
\begin {equation}
2\left( 1 + \frac{\beta}{\beta'}\right) \delta + 2\eta= dz \label{ghyp}
\end{equation}
where $\beta^{'}$ is defined for the active phase, eq.(11).

The derivation of (\ref{ghyp}) goes along the following lines .
It starts with eq.(\ref{rho}) for $\rho(r,t)$ and with the expressions
\begin{equation}
P(t)\sim t^{-\delta} \Phi(\Delta t^{1/{\nu_{\parallel}}})
\end{equation}
\begin{equation} 
P_{\infty} \sim \Delta^{\beta^{'}}, \, 
\beta^{'}=\delta \nu_{\parallel}
\end{equation}
for the survival probability. Since the stationary distribution is
unique
\begin{equation}
\rho(x,t) \rightarrow P_{\infty} \Delta^{\beta} \sim \Delta^{\beta + \beta^{'}}
\end{equation}
as $t\rightarrow \infty$. Hence $F(0,y)\sim y^{\beta + \beta^{'}}$ which
entails eq.(\ref{ghyp}).

In case of the DP transition (along the DP line of Fig.2) $\beta^{'} =
\beta$ as it is well known, and thus eq.(\ref{ghyp}) gives
\begin{equation}
2\delta+\eta=z/2
\label{DP}
\end{equation}

For the N-BARW2 transition, however, eq.(\ref{ghyp})
does not apply as eq.(\ref{rho}), according to the previous section,
is not an appropriate starting point.

To deduce the hyperscaling law valid for this case
 there are several possible ways of arguing.
It is possible to enlarge the parameter space of our model: 
we can think of a third direction in the parameter space,
approaching from where  the transition  turns out to be of
first order. For this aim one can introduce a 'magnetic field'
into the system by changing the annihilation probability
as  $w(+;--) =1-h$.
In this direction $\beta_h =0$ and thus 
eq.(\ref{ghyp}) gives (d=1):
\begin{equation}
 \eta + \delta = z/2
\label{gh}
\end{equation}
This law is satisfied for all the clusters investigated, including those
at the MDG point.
Even for $p_{ex}\neq 0$ we can still think of each point as 
being a first order transition point with $\beta=0$ in the $h$-direction
and the same considerations apply as above.
Thus on the basis of the results presented  now,
the conclusion to be drawn is that hyperscaling is generically 
satisfied in the whole N-BARW2 phase of the NEKIMA model.

Looking at the problem from a different point of view, however,
it is really not necessary to introduce the above auxiliary magnetic
field. Namely, one can simply make the observation, that all the
clusters investigated on  the $p_{ex}=0$ line are compact
 and from this fact
Eq(\ref{gh}) follows for the  hyperscaling law \cite{DiTre}.

Eq.(\ref{gh}) is known as the hyperscaling law for compact clusters.
By definition $\delta + \eta$ is the exponent which characterizes
the average population in surviving trials and the radius of such a
cluster grows as $R_{t} \sim t^{z/2}$. $\delta + \eta=d\frac{z}{2}$ is
simply the scaling law for the volume of a d-dimensional sphere of
radius $R_{t}$ \cite{DiTre}.

As a matter of fact, the '-' clusters are compact with $\delta=0$
also at
$p_{ex}\neq 0$, and even in the DP-region of Fig.2. This is not
true, however,  for the '+' cluster in the DP-region, which 
follows normal DP-cluster behavior (see TABLE II)
with the corresponding DP hyperscaling law
 eq.(\ref{DP}) (It is worth noting that whenever 
 $\delta=0$ , and this fails  only for the $'+'$ cluster in
the DP-region,  the CDP and DP hyperscaling laws do not differ).


\section {REENTRANT PHASE DIAGRAM, CLUSTER MF CALCULATIONS}

In the original NEKIM model at $\tilde\delta\geq 0$
 no transition
occurs, while for negative values of this parameter PC transition
takes place. The spin  asymmetry of NEKIMA changes the character of the 
transition into DP and this appears also for $\tilde\delta \geq 0$ 
Here we have chosen for our simulations and for the cluster mean-field
approximation calculations a fixed  negative value of $\tilde\delta$. 
Our aim has
 been to explore some possible reminiscence of the PC transition.
At the chosen parameter values $\tilde\delta=-.565, \Gamma=.5$  in NEKIM
the PC transition occurs at $p_{ex}=.12$ Turning to NEKIMA, at the
same values of $\tilde\delta, \Gamma$ our simulations show that the transition
point (which is DP, of course) shifts to $p_{ex}=.51$. The absorbing
phase below this point is all '-'. For letting $p_{+}< p$
the DP-line starts tangentially upon 
increasing $p_{ex}$ from $0$ and exhibits a reentrant property.
It ends up at $p_{ex}=.51$ tangentially. The regression takes place
at $p_{ex}\approx .12$, see Fig.2, most probably a remnant
of the transition point of the corresponding PC transition.
This turning point, however, is also of DP character as can be expected.

\begin{figure}
\begin{center}
\epsfxsize=80mm
\centerline{\epsffile{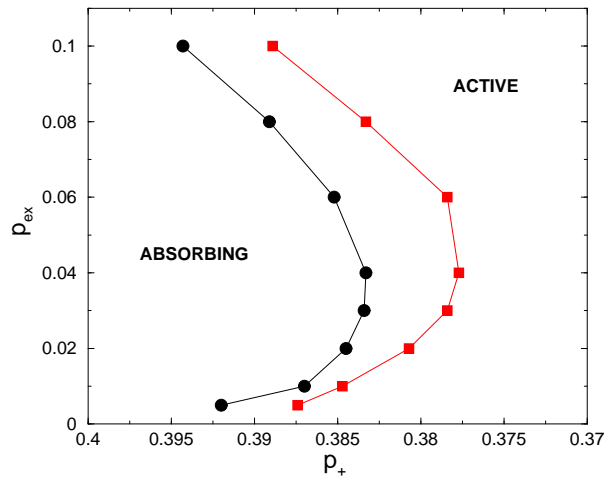}}
\caption{Steady state density in $N=6$ (bullets) and $N=7$ (boxes)
level approximation. Lines connecting symbols are shown for guidance 
of eye only.}
\label{GMF}
\end{center}
\end{figure}

Dynamical cluster mean-field approximations have been introduced for
nonequilibrium models by \cite{gut87,dic88}. The master equations for
$N=1-7$ block probabilities were set up
\begin{equation}
\frac{\partial P_N(\{s_i\})}{\partial t} = f\left (P_N(\{s_i\})\right) \ ,
\label{mastereq}
\end{equation}
where site variables may take values: $s_i=\pm 1$. 
Taking into account spatial reflection symmetries of $P_N(\{s_i\})$ this 
involves $72$ independent variables in case of $N=7$.
The equations were solved numerically for the
$\frac{\partial P_N(\{s_i\})}{\partial t} =0$ steady state condition,
for different $p_{ex}$ and $p_+$ values and the $\rho_k(\infty)$ kink 
density was expressed by $P_N(\{s_i\})$. The reentrant behavior could not be
observed for $N<6$ clusters. The results for $N=6,7$ are shown on
Fig. \ref{GMF}.
A slow shift towards lower $p_+$ values, which agrees with the simulations
can be observed.

 \section {DISCUSSION} 
We have investigated a one-dimensional nonequilibrium spin model
(NEKIMA) exhibiting strong spin-asymmetry. In the plane of two of the
parameters of NEKIMA (the kink-branching parameter and a spin-asymmetry
parameter) the phase diagram is as follows: besides a reentrant DP line
the NBARW-2 transition occurs at zero branching rate. Due to the
asymmetries, '+' and '-" spin clusters behave differently.
By investigating their  development we conclude that generic
power-law behavior characterizes the cluster
behavior at and in the vicinity of the NBARW-2 transition.

The critical cluster exponents obtained satisfy the
constraints on critical exponents in general: 1. $\delta \geq 0$ ,
2. $1 \leq z\leq 2$. The critical exponent $\delta$ has been found to be 
zero. The hyperscaling law is satisfied in the form known for
compact directed percolation, and indeed, the N-BARW2 
clusters are compact.

In a  different  problem Cafiero et al. \cite{Cafi}
have reported cluster exponents similar to the ones found
here. These authors studied how disorder affects the critical 
behaviour of DP-like  systems. 
Already in the eighties Noest \cite{noest} showed that quenched
disorder changes their behaviour in $d<4$ and demonstrated that when
$d=1$ a generic scale invariance can be observed. In \cite{Cafi}
it was shown that deep in the active phase $\eta=1$, $\delta=0$ and
$z=2$ for the model they considered. As we have found also 
generic scale invariance and the same exponents, in the active phase
of our model and even in the region which is the active phase of
the DP line of our phase diagram, the question arises whether
the similarity is fortuitous or not. Whether  the slowly diffusing '-'
clusters of NEKIMA distributed randomly in the x-direction
 can play a role similar to quenched impurities 
e.g. in the original NEKIM model is a question for 
future investigations.

\section{Acknowledgements}
Support from the Hungarian Research Fund OTKA (Grant Nos. T-025286 and
 T-034784) during this study is gratefully acknowledged. G. \'O. 
thanks the access to the
NIIFI Cluster-GRID and the Supercomputer Center of Hungary.

\end{document}